# Pressure-induced high-temperature superconductivity in ternary Y-Zr-H compounds

Wendi Zhao,[1] Defang Duan,[2,]* Hao Song,[1] Mingyang Du[2], Qiwen Jiang[2], Tiancheng Ma[2], Ming Xu[1], and Tian Cui[1,2,]*

[1]*Institute of High Pressure Physics, School of Physical Science and Technology, Ningbo University, Ningbo 315211, China*

[2]*State Key Laboratory of Superhard Materials, College of Physics, Jilin University, Changchun 130012, China*

*Corresponding author.*
*duandf@jlu.edu.cn*
*cuitian@nbu.edu.cn*

**Abstract:** Compressed hydrogen-rich compounds have received extensive attention as appealing contenders for superconductors, and further challenges are maintaining the stability and superconductivity of hydrides at lower pressures. In this work, we found several novel hydrides $YZrH_6$, $YZrH_8$ and $YZrH_{12}$ with excellent superconductivity in the Y-Zr-H ternary system. Interestingly, $YZrH_6$ with an A15-type structure can maintain dynamic stability down to 0.01 GPa and still with a critical temperature ($T_c$) of 16 K. $YZrH_8$ and $YZrH_{12}$ have high $T_c$ of 70 K and 183 K at 200 GPa and 160 GPa, respectively. The phonon modes associated with H atoms contribute significantly to the electron-phonon coupling, and the H-driven electronic density of states play an important role in superconductivity. These findings highlight relationship between the H-driven electronic density of states, electron-phonon coupling and the superconductivity in a distinct class of hydrides, opening new avenues for designing and optimizing new hydrogen-rich high temperature superconductors.

## 1. Introduction

The lightest and most abundant element in nature is hydrogen, and the insulated hydrogen molecules were predicted to be able to transform into atomic metallic phase at sufficient high pressure[1]. According to the Bardeen-Cooper-Schrieffer theory, atomic metallic hydrogen is considered as a potential room-temperature superconductor due to the high Debye temperature and strong electron-phonon coupling[1, 2]. Although there have been many attempts to realize metallic hydrogen (pressures up to 500 GPa), the related reports are still controversial[3-5]. In 2004, it is proposed that introducing other elements into pure hydrogen to form hydrides can promote the transformation of the hydrogen into the metallic phase at low pressure, known as the effect of "chemical pre-compression"[6]. At present, almost all binary hydrides have been intensively studied in theoretical and a considerable part has been experimentally confirmed. For example, covalent hydride $H_3S$ and clathrate hydrides $CaH_6$, $YH_6$, $LaH_{10}$, etc.[7-13] all exhibit superconducting transition temperatures above 200 K. As is well known, the clathrate hydrides widely exist in alkaline earth and rare earth metal hydrides, and clathrate hydrides $CaH_6$ containing $H_{24}$ cage was predicted with remarkably high $T_c$ of 220-235 K at 150 GPa[11]. Its high $T_c$ is attributed to the physical characteristics and properties of the clathrate hydrogen lattice close to the metallic hydrogen, and the Ca atoms as electron donors play an important role in stabilizing the $H_{24}$ cage. Recently, $CaH_6$ has been successfully synthesized and the measured superconducting critical temperature $T_c$ is 215 K at 172 GPa[14], which is in good agreement with the previous calculation. Furthermore, some hydrides with A15-type structures are also predicted to have high $T_c$, such as $AlH_3$, $ZrH_3$, $MgSiH_6$, $LiPH_6$, etc. [15-20], among which $LiPH_6$ is predicted to exhibit a high $T_c$ of 167 K at 200 GPa[18].

The fruitful results of searching for high-temperature superconductors in binary hydrides have promoted the relevant research of ternary hydrides which have more abundant stoichiometries and structures. Incorporating new elements directly into well studied binary hydrides is an effective way to construct ternary hydrides. The lithium-doped magnesium hydride $MgH_{16}$, which contains a lot of $H_2$ molecular units, and the introduction of extra electrons effectively promoted the dissociation of $H_2$ molecular units, further forming $Li_2MgH_{16}$ with remarkably high estimated $T_c$ of ~473 K at 250 GPa[21]. Another optimized strategy is engineering binary hydride backbones, which are easier to metallize than pure H backbones and can be designed by doping known structures with additional atoms, which break the local motif of the parent structure. For example, Zhang et al. theoretically predict that $LaBeH_8$ with a "fluorite-type" H-Be alloy backbones can maintain dynamic stability as low as 20 GPa and exhibit a high $T_c$ of 185 K[22]. Notably, the superconductivity of some binary hydrides is well preserved by doping with Y element, such as ternary clathrate hydrides $YCaH_{12}$, $LaYH_{12}$, $YLuH_{12}$, etc[23-27]. Recently, a series of La-Y-H ternary compounds have been successfully synthesized recently[23]. As is well known, Y and Zr atoms have similar electronegativity and atomic radius. High temperature superconductors are very likely to exist in ternary Y-Zr-H system.

We performed the systematical structure searches in the Y-Zr-H system under high pressure and discovered several novel hydrides: stable $Pm\overline{3}$-$YZrH_6$, $P4/mmm$-$YZrH_8$ and metastable $Pm\overline{3}m$-$YZrH_{12}$. $Pm\overline{3}$-$YZrH_6$ has an A15-type structure. Both $P4/mmm$-$YZrH_8$ and $Pm\overline{3}m$-$YZrH_{12}$ contain clathrate hydrogen sublattices. Electron-phonon coupling (EPC) calculations indicate that they are both potential high-temperature superconductors. Especially $YZrH_6$ can maintain dynamic stability down to 0.01 GPa, which is favorable for synthesis and applications at ambient pressure.

2. **Computational methods**

We performed the high-pressure structure searches of $YZrH_x$ (1≤x≤ 12) with 1–2 formula units using AIRSS (Ab initio Random Structure Searching) code[28, 29] at 200 GPa. A plane-wave cut-off of 500 eV and Monkhorst-Pack meshes $k$-point spacing of 2π×0.07 Å$^{-1}$ were used. All predicted structures have been re-optimized by the on-the-fly (OTF) generation of ultrasoft pseudopotentials in CASTEP (Cambridge Sequential Total Energy Package) code[30], where the valence electrons configurations are $4s^24p^64d^15s^2$ for Y, $4s^24p^64d^25s^2$ for Zr and $1s^1$ for H. The Brillouin zone was sampled with a $k$-point mesh of $2\pi \times 0.03$ Å$^{-1}$, and the cut-off energy of 800 eV was chosen to ensure that the convergence of enthalpy within 1 meV/atom. Electronic properties were calculated within the framework of density functional theory as implemented in Vienna ab initio simulation package (VASP)[31]. The Monkhorst-Pack *k*-mesh with grid spacing of $2\pi \times 0.02$ Å$^{-1}$ and energy cut-off of 1000 eV were adopted. The exchange-correlation functional was described using the Perdew-Burke-Ernzerhof (PBE) parametrization within the generalized gradient approximation (GGA). The ion-electron interactions part was implemented with the projector augmented wave (PAW) method[32]. The EPC matrix elements were calculated using density functional perturbation theory as implemented in the Quantum ESPRESSO package[33]. The ultrasoft pseudopotential were selected with a kinetic energy cut-off of 80 Ry. A 24×24×24 *k*-point grid and a 6×6×6 *q*-point grid were chosen for YZrH$_6$ and YZrH$_{12}$. A 15×15×12 *k*-point grid and a 5×5×4 *q*-point grid were chosen for YZrH$_8$. The Allen-Dynes modified McMillan equation[34] and Gor'kov-Kresin equation (G-K)[35] were used to calculate the superconducting critical temperature.

## 3. Results and discussion

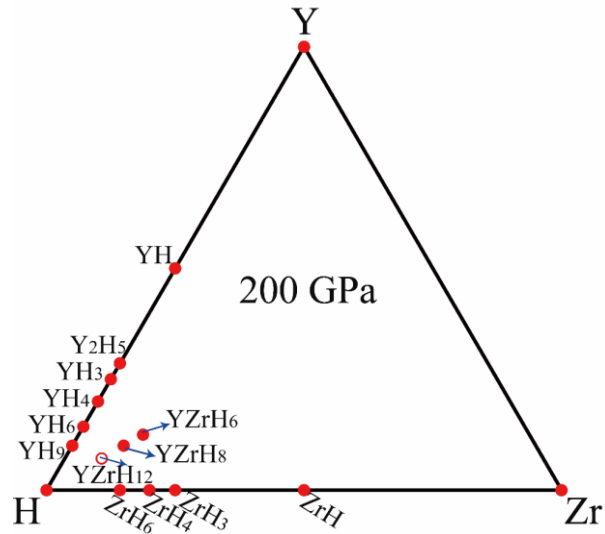

Fig.1 The ternary phase diagram (convex hull) of the Y-Zr-H system at 200 GPa. The red solid circles indicate stable structures, and the red hollow circles indicate metastable structures. The structures of stable element (Y, Zr and H) and binary compounds locating on the edges of the triangle were obtained from previous work[15, 36-40].

We performed structure searches in the ternary YZrH$_x$ (1≤x≤12) system at 200 GPa. The thermodynamic stability of all predicted high-pressure structures was determined by constructing a ternary convex hull (see Fig.1). The results show that YZrH$_6$ and YZrH$_8$ fall on the convex hull, while the YZrH$_{12}$ is 45 meV above the convex hull at 200 GPa, which is thermodynamic metastable. Considering that many metastable compounds have been synthesized experimentally in the inorganic crystal database[41, 42], YZrH$_{12}$ still has the possibility of being synthesized in high-pressure experiments. As for YZrH$_6$, there are two competitive phases $R\bar{3}$ and $Pm\bar{3}$, which have very close enthalpies (see Fig. 2a). Since highly symmetric structures tends to possess high-temperature superconductivity, the cubic $Pm\bar{3}$ phase will be the focus of our discussion.

We analyzed possible high-pressure synthesis routes for YZrH$_6$, YZrH$_8$ and YZrH$_{12}$. It is reported that a series of Y-H compounds were synthesized using YH$_3$+H$_2$ or Y+H$_2$ as reactants[40], $Pm\bar{3}n$-ZrH$_3$ was synthesized using ZrH$_2$+H$_2$ or Zr+H$_2$ as reactants, and it remained stable above 10 GPa[15]. These results show that binary hydrides YH$_3$, ZrH$_3$, ZrH$_2$ and pure elements Y, Zr, H$_2$ are possible reactants. As shown in Fig. 2, we mainly calculated the formation enthalpies of YZrH$_6$, YZrH$_8$ and YZrH$_{12}$ relative to these reactants. YZrH$_6$ has a lower enthalpy of formation relative to multiple synthetic routes. Especially at ambient pressure, its significant energetic advantage over synthetic routes (Y+Zr+H$_2$ and Y+ZrH$_2$+H$_2$, etc.), means it is promising to be synthesized at ambient pressure. Relative to the different synthetic routes, YZrH$_8$ and YZrH$_{12}$ have the lowest enthalpies above 140 GPa and 180 GPa, respectively, suggesting that they can be synthesized at high pressure. We further confirmed the dynamic stability of YZrH$_6$, YZrH$_8$ and YZrH$_{12}$ by calculating the phonon dispersion curves (see Fig. 6b and Fig. 7). The minimum stable pressures of YZrH$_8$ and YZrH$_{12}$ are 200 GPa and 160 GPa, respectively. Surprisingly, YZrH$_6$ can maintain dynamic stable down to 0.01 GPa. In the following discussion, we will focus on their structure and properties at the lowest stable pressure.

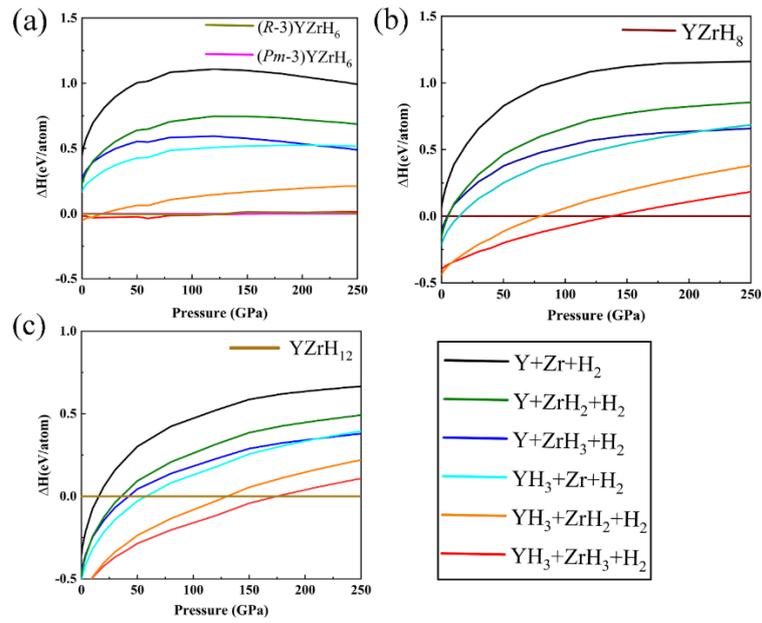

Fig. 2 The enthalpy difference curves with respect to (a) YZrH$_6$, (b) YZrH$_8$ and (c) YZrH$_{12}$ from divergent synthetic routes as a function of pressure.

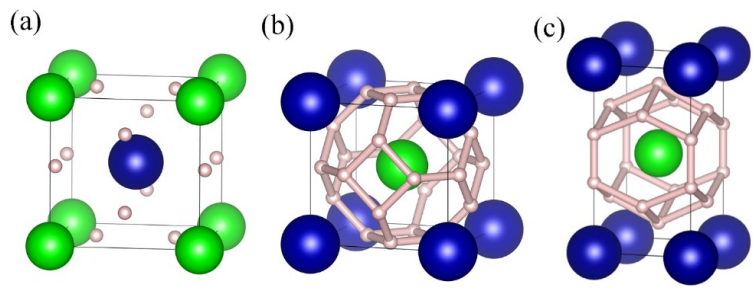

Fig.3 The crystal structure of (a) YZrH$_6$ at 0.01 GPa, (b) YZrH$_{12}$ at 160 GPa and (c) YZrH$_8$ at 200 GPa. The blue, green, and pink spheres represent Y, Zr and H atoms, respectively.

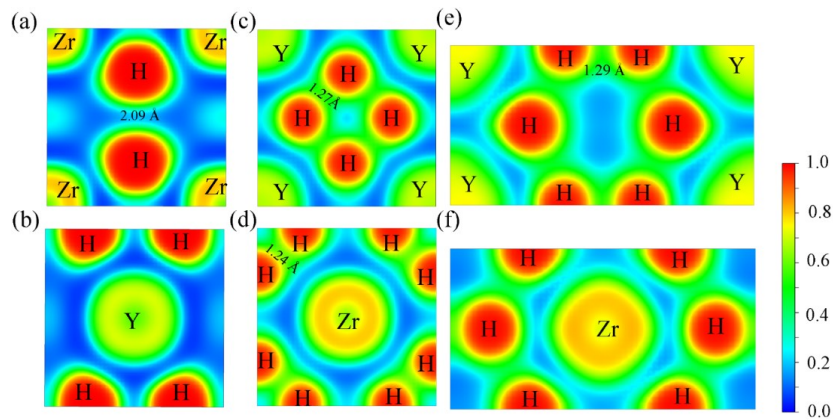

Fig.4 The electron localization functions of Y-H plane and Zr-H plane in (a-b) YZrH$_6$ at 0.01 GPa, (c-d) YZrH$_{12}$ at 160 GPa and (e-f) YZrH$_8$ at 200 GPa.

The structure of $Pm\bar{3}$-YZrH$_6$ is very similar to that of A15-type hydrides (AlH$_3$, GaH$_3$, ZrH$_3$, etc[15, 19, 20].), the difference is that its metal atom positions are inequivalent, as shown in Fig. 3. For $Pm\bar{3}$-YZrH$_6$, the Y and Zr occupy 1a (0.000, 0.000, 0.000) and 1b (0.500, 0.500, 0.500) sites, respectively, forming a cubic lattice. All equivalent H atoms occupy 6f (0.237, 0.000, 0.500) sites. The nearest H-H distance is 2.09 Å at 0.01 GPa, which is much longer than the bonding distance of H-H bond. As the case of $Pm\bar{3}m$-YZrH$_{12}$, metal atoms Y and Zr also form a cubic lattice. The H atoms form the H$_{24}$ cage around the metal atoms, which consists of six quadrilaterals and eight hexagons. The H-H distance is 1.24 Å-1.27 Å at 160 GPa. Similar clathrate structures also appear in the previously reported ternary hydrides YCaH$_{12}$, LaYH$_{12}$, etc[23-25]. For $P4/mmm$-YZrH$_8$, H atoms form H$_{18}$ cages around metal atoms, the H-H distance range is 1.29-1.49 Å at 200 GPa. Interestingly, YZrH$_8$ can also be obtained by substitution of binary hydrides $I4/mmm$-YH$_4$ or $I4/mmm$-ZrH$_4$[39, 43, 44]. This indicates that elements with similar atomic radius and electronegativity may share the same hydrogen sublattice and further form new ternary hydrides.

The calculated electron localization function (ELF) and bader charge were used to analyze the electronic properties of YZrH$_6$, YZrH$_{12}$ and YZrH$_8$. The common feature is the low ELF values between the metal atoms and H, which confirm the existence of ionic bonds (see Fig. 4). It is worth noting that there is a low ELF value between the nearest H atoms in YZrH$_6$, which is consistent with the long H-H bond length, reflecting the ionic character of the structure. Bader charge analysis of YZrH$_6$ at 0.01 GPa shows that each Y and Zr atom loses charges of 1.73 and 1.69 |e|, respectively, and each H atom gains 0.57 |e| charges. Furthermore, the ELF values between the nearest-neighbor H-H in YZrH$_{12}$ and YZrH$_8$ are 0.6 and 0.5, respectively, indicating the formation of weak covalent bonds (see Fig. 4). Bader charge analysis reveals that each Y and Zr atom in YZrH$_{12}$ at 160 GPa transfer 1.39 and 1.47 |e| charges to the H atom, respectively, so that each H atom gains an average of 0.24 |e| charges. For YZrH$_8$, each Y and Zr atom loses 1.27 and 1.46 |e| charges at 200 GPa, respectively, and each H atom gains 0.34 |e| charges. Obviously, the Y and Zr atoms in YZrH$_6$, YZrH$_{12}$ and YZrH$_8$ are good electron donors, and the H-H bonds are elongated due to the extra electron occupying its anti-bonding states. Notably, each H atom in YZrH$_6$ obtains the most electrons from Y and Zr, so it can remain stable in the form of atomization. In addition, there are weak covalent interactions between the H atoms in both YZrH$_{12}$ and YZrH$_8$, which are beneficial to enhance the hydrogen-derived electronic density of states at the Fermi level.

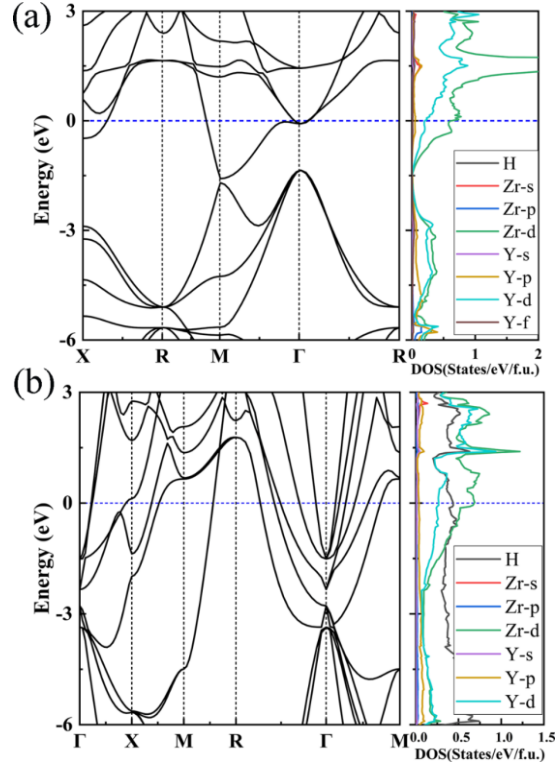

Fig. 5 Calculated electronic band structures and projected density of states (PDOS) for (a) YZrH$_6$ at 0.01 GPa (b) YZrH$_{12}$ at 160 GPa. The Fermi level is set to zero.

The calculated electronic band structures and projected density of states of YZrH$_6$ and YZrH$_{12}$ are shown in Fig. 5. The overlapping of valence and conduction bands indicates that they are metallic. For YZrH$_6$, there are two obvious flat bands above the Fermi level about 1.5 eV, which are caused by the local $d$ electrons of Y and Zr. They also further dominate the electronic density of states at the Fermi level. As the case of YZrH$_{12}$, many steep conduction bands cross the Fermi level around the Γ point and form multiple deep "electron pockets". Note that high H-derived electronic density of states, which appearing near the Fermi level, is beneficial to strong electron-phonon coupling. As for YZrH$_8$ (see Fig. 6a), electron pockets near the Fermi level form around the Γ and R points while H has a significant contribution to the density of electronic states at the Fermi level, these properties imply the potential superconductivity of the system.

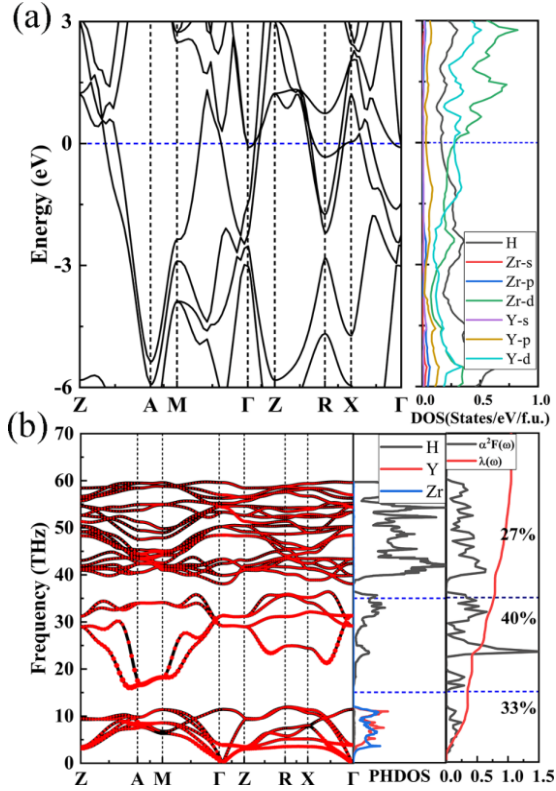

Fig. 6 (a) The electronic band structures and atom-projected density of states for $P4/mmm$-YZrH$_8$ at 200 GPa. (b) Phonon dispersion curves, projected phonon density of states (PHDOS), and Eliashberg spectral function $\alpha^2F(\omega)$ together with the electron-phonon integral $\lambda(\omega)$ for $P4/mmm$-YZrH$_8$ at 200 GPa. The size of the red solid dots in the phonon spectra is proportional to the strength of electron-phonon coupling.

We further calculated the phonon dispersion curves, projected phonon density of states (PHDOS), Eliashberg spectral functions $\alpha^2F(\omega)$ with the electron-phonon integral $\lambda(\omega)$ of YZrH$_8$, YZrH$_{12}$ and YZrH$_6$, as shown in Fig. 6b and Fig. 7. The phonon modes in the low frequency region are mainly associated with the Y and Zr atoms due to their heavy atomic masses. The lighter H atoms drive the phonon modes in the mid-high frequency region. Note that the high peaks of the Eliashberg spectral function $\alpha^2F(\omega)$ appear in the middle frequency region, which are mainly caused by the soft phonon modes in this region. The soft phonon modes of YZrH$_8$ are distributed between 15 and 35 THz, contributing up to 40 % of the total $\lambda$ (see Fig. 6b). For YZrH$_6$, the contribution of the soft phonon modes between 10 and 25 THz is also approximately 40 % (see Fig. 7a). At the same time, the phonon modes in the low-frequency region (< 10 THz) mainly related to Y and Zr atoms also have a significant contribution to the total $\lambda$ up to 47 %. Interestingly, an appealing soft phonon mode in YZrH$_{12}$ appears at the M point (see Fig. 7b), resulting in phonon modes in the mid-frequency (10-30 THz) region contributing as high as 48 % to the total $\lambda$. We use the size of the red dots on the phonon spectrum to represent the contribution of different phonon modes to the electron-phonon coupling, and find that the dense red dots are decorated on the soft phonon modes, indicating that they effectively enhance the electron-phonon coupling.

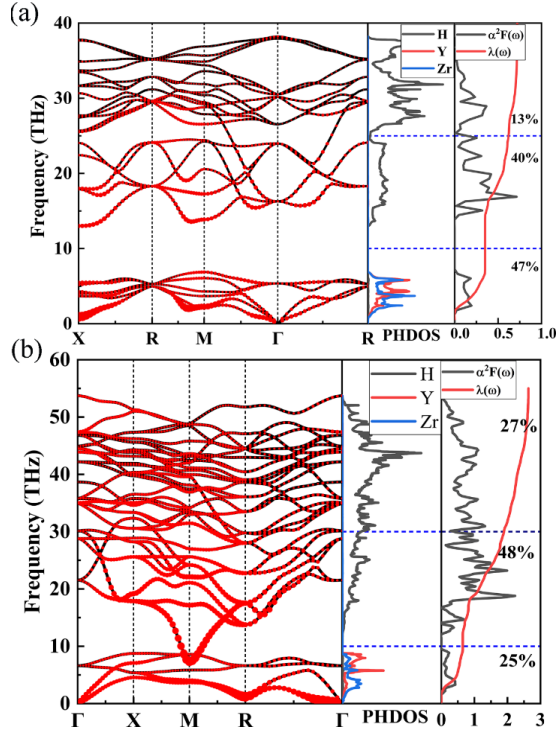

Fig.7 Calculated phonon dispersion curves, projected phonon density of states (PHDOS), and Eliashberg spectral function $a^2F(\omega)$ together with the electron-phonon integral $\lambda(\omega)$ for (a) $Pm\bar{3}$-YZrH$_6$ at 0.01 GPa (b) $Pm\bar{3}m$-YZrH$_{12}$ at 160 GPa. The size of the red solid dots in the phonon spectra is proportional to the strength of electron-phonon coupling.

We evaluated the superconducting critical temperatures of YZrH$_8$, YZrH$_{12}$ and YZrH$_6$ by solving the Allen-Dynes modified McMillan equation:[34]

$$T_c = \omega_{log} \frac{f_1 f_2}{1.2} \exp\left(\frac{-1.04(1+\lambda)}{\lambda - \mu^* - 0.62\lambda\mu^*}\right) \quad (1)$$

where $f_1$ and $f_2$ are two correction factors. The Coulomb pseudopotential $\mu^*$ is set to the typical 0.1-0.13. The logarithmic average phonon frequency $\omega_{log}$ and EPC parameter $\lambda$ were given by:

$$\omega_{log} = \exp\left(\frac{2}{\lambda}\int\frac{d\omega}{\omega}\alpha^2 F(\omega)\ln(\omega)\right) \qquad (2)$$

$$\lambda = 2\int\frac{\alpha^2 F(\omega)}{\omega}d\omega \qquad (3)$$

The calculated $T_c$ of YZrH$_8$ at 200 GPa is 70-60 K. The $\lambda$ and $\omega_{log}$ are 1.05 and 866.54 K, respectively. YZrH$_{12}$ exhibits the highest $T_c$ 183-167 K at 160 GPa with $\lambda$ and $\omega_{log}$ of 2.63 and 803 K, respectively. To evaluate the contribution of the optical branch of the phonon spectrum to the electron-phonon coupling, we calculated the $T_c$ of YZrH$_{12}$ using the G-K equation[35], as shown in Fig.8a. The coupling constant $\lambda_{opt}$ in the G-K equation is used to describe the interaction between electrons and optical phonons. The $T_c$ obtained by solving the G-K equation at 160 GPa is 207 K. The increase in pressure reduces the density of states at the Fermi surface, thereby weakening the interaction of electrons with optical phonons, so the coupling constant $\lambda_{opt}$ gradually decreases. The $T_c$ value slightly decreases. In addition, the phonon frequency increases with increasing pressure, resulting in a larger $\omega_{log}$. Therefore, there is no significant fluctuation of $T_c$ under the competitive effect of $\omega_{log}$ and $\lambda$.

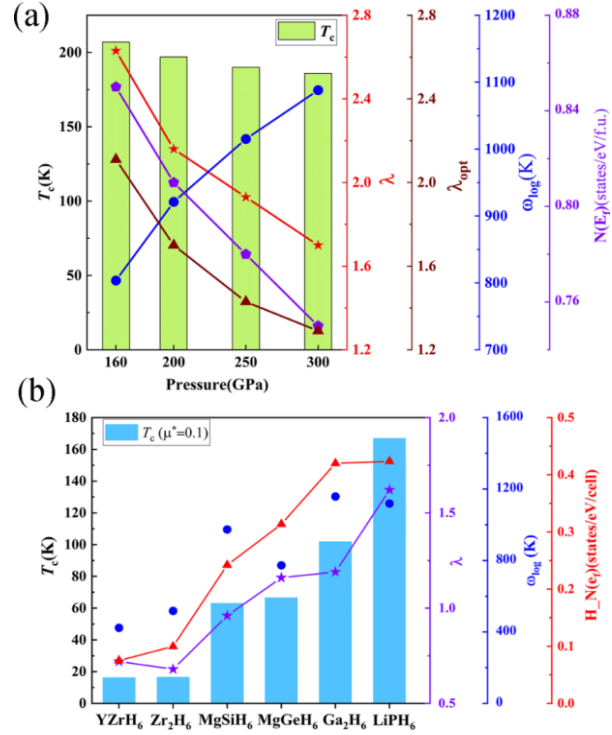

Fig. 8 (a) Calculated superconducting parameters of YZrH$_{12}$ at different pressures. The obtained $T_c$ using GK equation ($\mu^* = 0.1$), strength of the interaction of electrons with optical phonons ($\lambda_{opt}$), electronic density of states at the Fermi energy ($N_{\varepsilon f}$). (b) The $T_c$ and related parameters of A15 hydrides. The superconducting parameters of ZrH$_3$, MgSiH$_6$, MgGeH$_6$, GaH$_3$ and LiPH$_6$ were obtained from previous work[15-19]. The $T_c$s are calculated using the Allen-Dynes modified McMillan equation ($\mu^* = 0.1$), and the DOS of hydrogen at the Fermi energy (H\_$N_{\varepsilon f}$).

The EPC constant $\lambda$ and logarithmic average phonon frequency $\omega_{log}$ of YZrH$_6$ at 0.01 GPa are 0.72 and 423 K, respectively. Its $T_c$ value reaches 16-13 K (see Fig. 8b). We further summarize the predicted $T_c$ and related parameters of several A15-type hydrides in order to better explore the superconductivity of these hydrides, as shown in Fig. 8b. Obviously, the predicted $T_c$ values generally increase with the density of states of hydrogen at the Fermi level and follow the same trend as electron-phonon coupling $\lambda$. H atoms drive the higher density of electronic states at the Fermi level and dominate phonon density of states, resulting in strong electron-phonon coupling. LiPH$_6$ not only has strong electron-phonon coupling, but also exhibits higher H\_$N_{ef}$, and thus has highest $T_c$. Ternary A15-type hydrides are very abundant and have not been fully explored. Choosing suitable elements to induce hydrogen-derived high electron density of states at the Fermi level is an effective strategy to obtain excellent superconductivity. Although YZrH$_6$ has a lower $T_c$, it remains stable at near-ambient pressures, which is beneficial for both experimental synthesis and practical applications.

## 4. Conclusions

In summary, we systematically investigated the structure and properties of ternary Y-Zr-H compounds under high pressure, and found stable YZrH$_6$, YZrH$_8$ and metastable YZrH$_{12}$. YZrH$_6$ has an A15-type structure and maintain dynamic stability down to 0.01 GPa with $T_c$ of 16 K. For this class of A15-type hydrides, the high contribution of hydrogen to the electronic density of states at the Fermi level is the key to induce high $T_c$. YZrH$_8$ and YZrH$_{12}$ with clathrate hydrogen sublattices exhibit high superconducting transition temperatures of 70 K and 183 K at 200 GPa and 160 GPa, respectively. The soft phonon modes associated with the vibrations of hydrogen atoms in all three hydrides play an important role in enhancing the electron-phonon coupling. Our comprehensive analysis of the stability and superconductivity of Y-Zr-H compounds will motivate further theoretical and experimental studies.

**Conflicts of interest**

The authors declare no competing financial interest.

**Acknowledgements**

This work was supported by the National Key R&D Program of China (No. 2018YFA0305900), and the National Natural Science Foundation of China (Grants No. 52072188, No. 12122405, No. 51632002). Program for Changjiang Scholars and Innovative Research Team in University (Grant No. IRT_15R23), Some of the calculations were performed at the High Performance Computing Center of Jilin University and using TianHe-1(A) at the National Supercomputer Center in Tianjin.